\pdfoutput=1
\documentclass[letterpaper, 12pt]{article}
\usepackage{siunitx,wrapfig,amsmath}
\usepackage[utf8]{inputenc}
\usepackage[english]{babel}
\usepackage{graphicx, float} 
\usepackage{fancyhdr}
\usepackage[sfdefault]{roboto}
\usepackage[T1]{fontenc}
\usepackage{xcolor}
\usepackage{sectsty}
\usepackage{enumitem}
\usepackage{listings}
\usepackage{subcaption}

\definecolor{BaseDark}{HTML}{114B5F}
\definecolor{BaseLight}{HTML}{028090}
\definecolor{TextGray}{HTML}{465463}

\sectionfont{\color{BaseLight}}
\subsectionfont{\color{BaseDark}}
\paragraphfont{\normalfont \color{TextGray}}

\makeatletter
\renewcommand{\@seccntformat}[1]{}
\makeatother

\lstdefinestyle{stan}{
	language=Java,
	basicstyle={\scriptsize\ttfamily},
	commentstyle=\color{gray},
	numbers=left,
	numberstyle=\tiny\color{gray},
	numbersep=5pt,
	breaklines=true
}

\title{Large‐Scale Statistical Survey of Magnetopause Reconnection \\ 
	\ \\
	\large Computer Science\\
	University of New Hampshire\\
	Durham, New Hampshire, USA}
\author{Samantha Piatt}
\date{May 7, 2019}

\begin{document}
	\pagestyle{fancy}
	
	\maketitle
	
	\thispagestyle{empty}
	
	\section{Abstract}
	The Magnetospheric Multiscale Mission (MMS) seeks to study the micro-physics of reconnection, which occurs at the magnetopause boundary layer between the magnetosphere of Earth and the interplanetary magnetic field originating from the sun. Identifying this region of space automatically will allow for statistical analysis of reconnection events. The magnetopause region is difficult to identify automatically using simple models, and time consuming for scientists to classify by hand. We introduced a hierarchical Bayesian mixture model with linear and auto regressive components to identify the magnetopause. Using data from the MMS mission with the programming languages R and Stan, we modeled and predicted possible regions and evaluated our performance against a boosted regression tree model. Our model selects twice as many magnetopause regions as the comparison model, without significant over selection, achieving a 31\% true positive rate and 93\% true negative rate. Our method will allow scientists to study the micro-physics of reconnection events in the magnetopause using the large body of MMS data without manual classification.
	
	\section{Introduction}
	\begin{figure}
		\centering
		\includegraphics[width=0.7\linewidth]{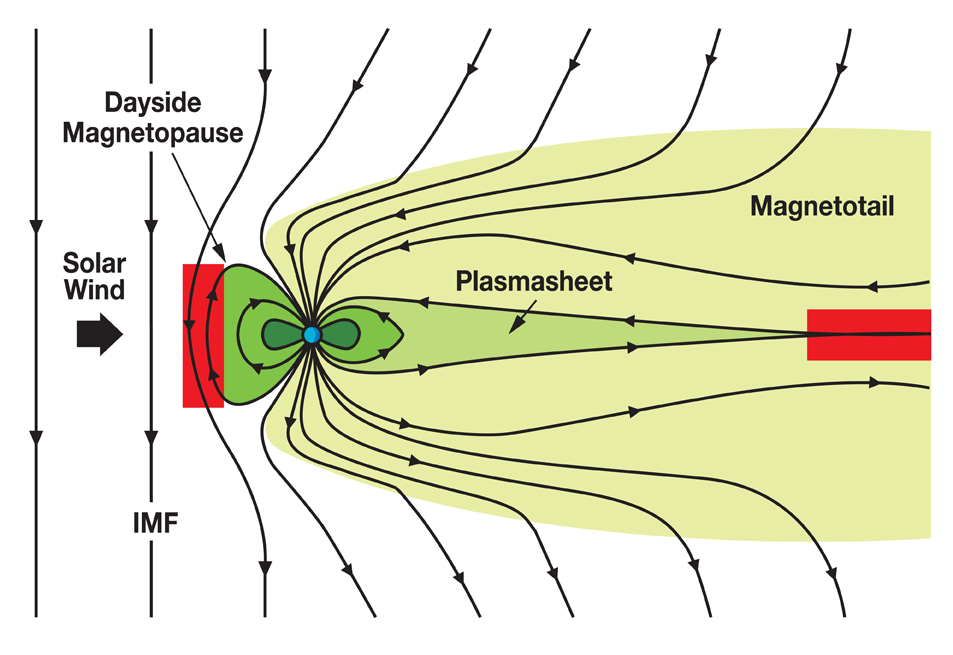}
		\caption{Magnetic fields.}
		\label{fig:megnetopause}
	\end{figure}
	Magnetic reconnection is an important phenomenon in physics, which happens on a large and small scale. The Magnetospheric Multiscale (MMS) project aims to study the micro-physics of magnetic reconnection. It consists of four satellites orbiting the Earth in a tetrahedral formation, capturing 3-dimensional data with multiple instruments to record magnetic field, ion, and electron data.
	
	Because the micro physics of reconnection are hard to identify in low resolution data, MP regions (which are easier to characterize), are selected instead. During the selection process, a scientist-in-the-loop (SITL) verifies the selections made by on-board algorithms in low-resolution data, confirming or adjusting selected regions that they believe are of interest \cite{Fuselier2016}. The selections are then transmitted to Earth in high-resolution, depending on bandwidth available. Each selection also receives a priority code as well as a comment to note what the SITL believed made this time region significant.
	
	A set of guidelines make this process consistent and standard between different scientists, but variations from subjectivity exist. Additionally, selections are subject to bandwidth constraints, and therefore SITLs take this into account when prioritizing which data are to be transmitted in high resolution. The manpower and subjectivity inherent in selecting interesting regions present a problem that can be solved with machine learning.
	
	We generated a model that gives any scientist looking to study reconnection, or the magnetopause, a more automated and statistically meaningful way of selecting these important regions of interest. Reconnection events happen in either the dayside magnetopause (MP) or the night-side magentotail – herein, we focus on the MP events only. Our model attempts to identify the MP boundary layer, an arbitrary boundary between the Magnetosphere (Earth’s magnetic field) and the magnetosheath (solar wind facing side of the boundary layer). Each satellite captured data relevant to the MP during phase 1 of their mission, March 2015 through February 2017.
	
	\section{Methods}
	Data for training and testing was sourced from only one of the four satellites during a period of 2 months from January through February 2017. Specifically, the data consists of readings from the Dual Ion Spectrometer (DIS) and Dual Electron Spectrometer (DES) \cite{Pollock2016}, as well as the Fluxgate Magnetometer (FGM) \cite{Russell2016, Torbert2016}, which together record the magnetic field and ion density, temperature, and pressure. Data from the slow survey is interpolated to that of the DES at 4.5 second intervals. This subset of data contains 395,458 rows, with 246 excluded due to missing values.
	
	\subsection{Modeling in Stan and R}
	The raw data is parsed by removing empty elements, grouping by effective orbit, and then split into test and training sets. Data was pre-processed in R by segmenting different orbits by identifying gaps in timestamp values greater than 100 seconds. 14 orbits were selected randomly, with 80\% of those orbits used for training data and 20\% used for test data. This results in 119,910 data points for training and 33,217 data points for testing, totaling 153,127 rows of data.
	
	We use several values in the low-resolution data from the DIS and FGM instruments in our model. The raw data is parsed by removing empty elements and grouping by effective orbit. We generate the $\theta_{CA}$ and $T_i$ as in equations \ref{eq:process:t} and \ref{eq:process:ca}, and then transform $n_i$ and $T_i$ by taking their natural log values.
	
	\begin{align}
	\label{eq:process:t}
	T_i = & \frac{1}{3} \left(T_{i,\parallel} + 2 T_{i,\bot} \right) \\
	\label{eq:process:ca}
	\theta_{CA} = & \tan^{-1}{B_{y} / B_{z}}
	\end{align}
	
	Our model was written in Stan for both training and testing phases. Stan is both a probabilistic programming language and a sampler that does Bayesian inference. It uses variations of gradient descent for sampling and optimization, and can run on command line, as well as through R or Python \cite{StanManual}. For our model, 100 iterations were used for both training and testing, on one core. Training and testing model code is included in the appendix, Listings \ref{code:training} and \ref{code:testing}.
	
	\subsection{Mixture Model}
	\begin{figure}
		\centering
		\begin{subfigure}[b]{0.45\textwidth}
			\includegraphics[width=\textwidth]{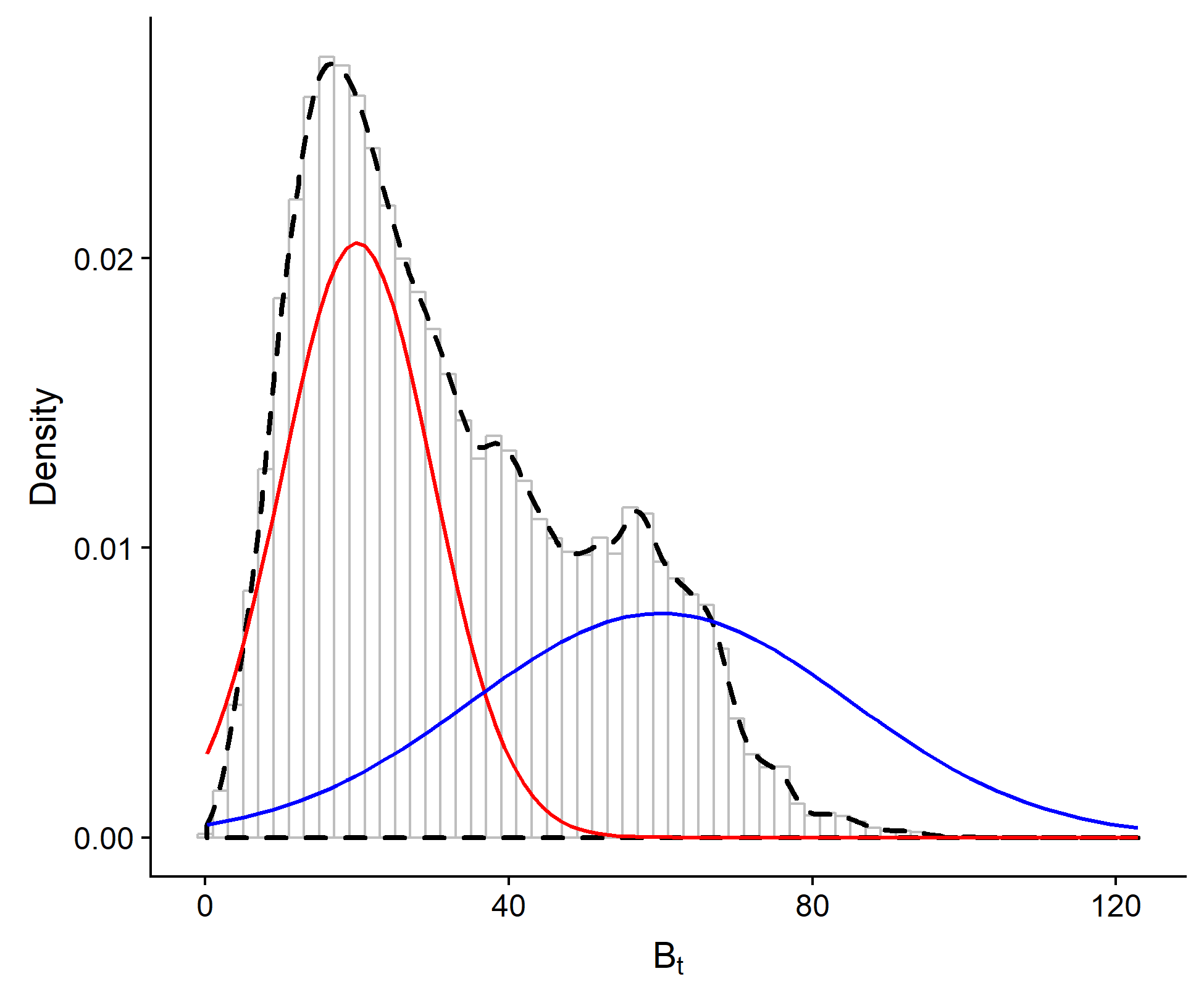}
			\caption{}
			\label{fig:dist.bt}
		\end{subfigure}
		\begin{subfigure}[b]{0.45\textwidth}
			\includegraphics[width=\textwidth]{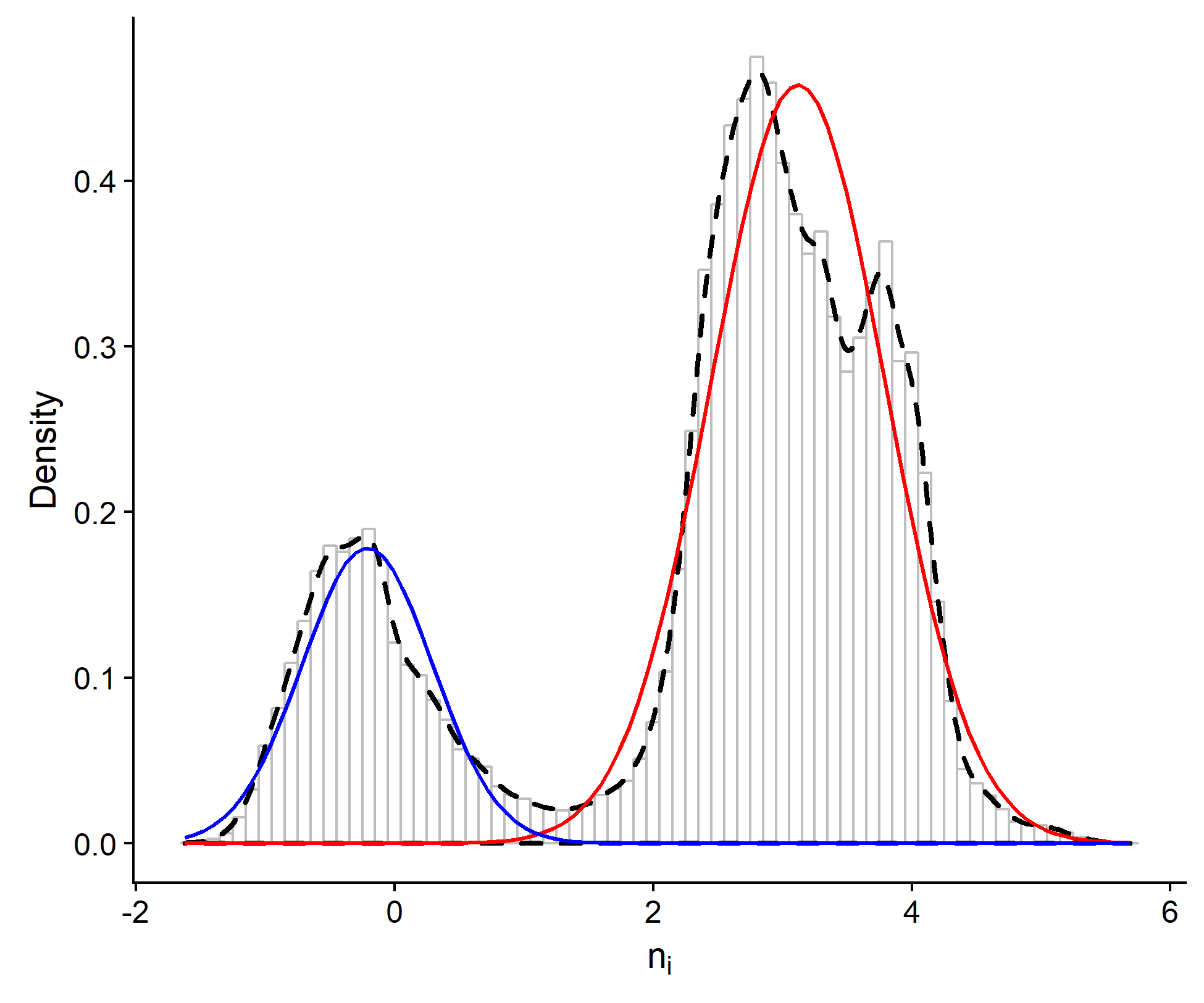}
			\caption{}
			\label{fig:dist.n}
		\end{subfigure}
		\begin{subfigure}[b]{0.45\textwidth}
			\includegraphics[width=\textwidth]{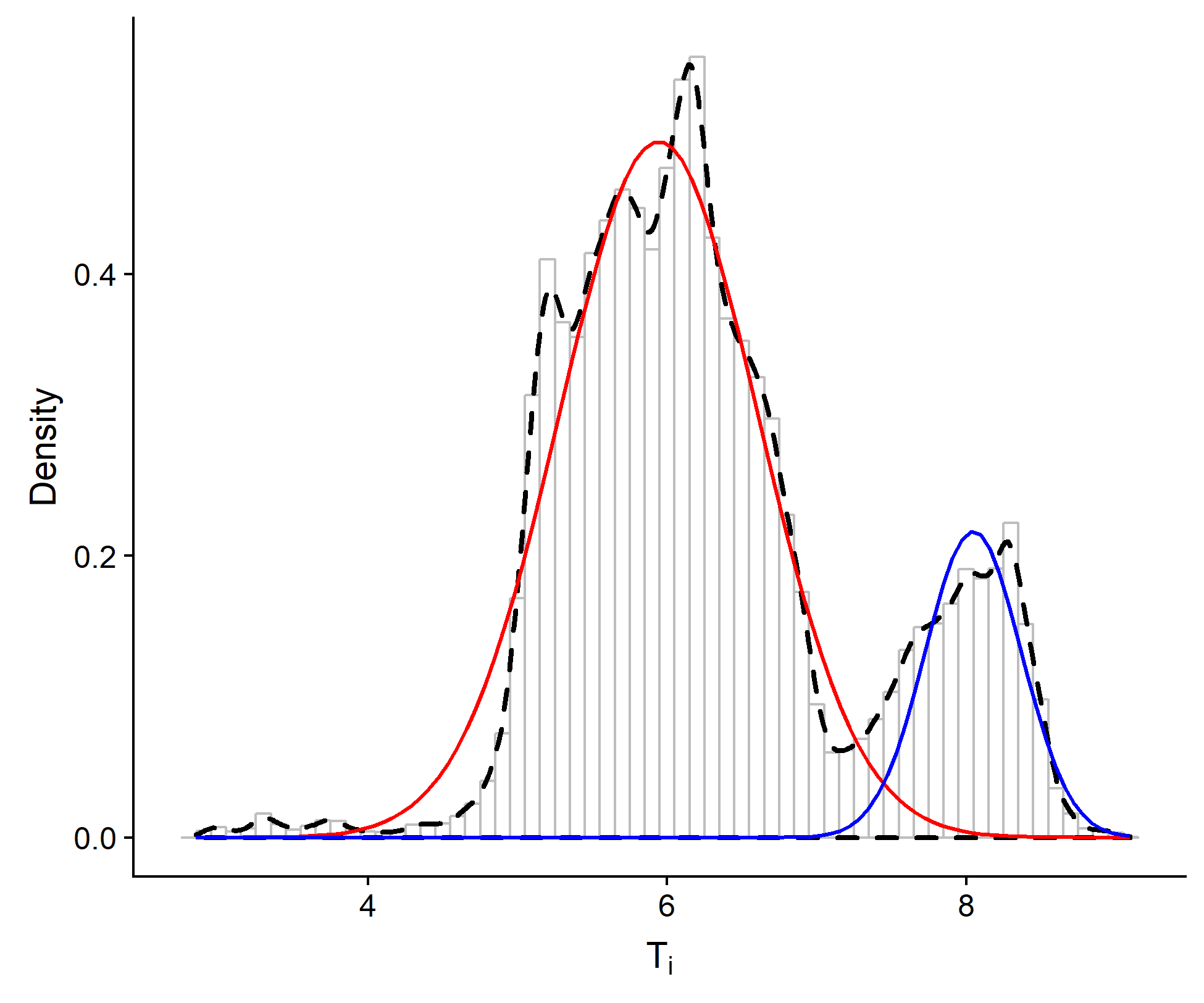}
			\caption{}
			\label{fig:dist.t}
		\end{subfigure}
		\begin{subfigure}[b]{0.45\textwidth}
			\includegraphics[width=\textwidth]{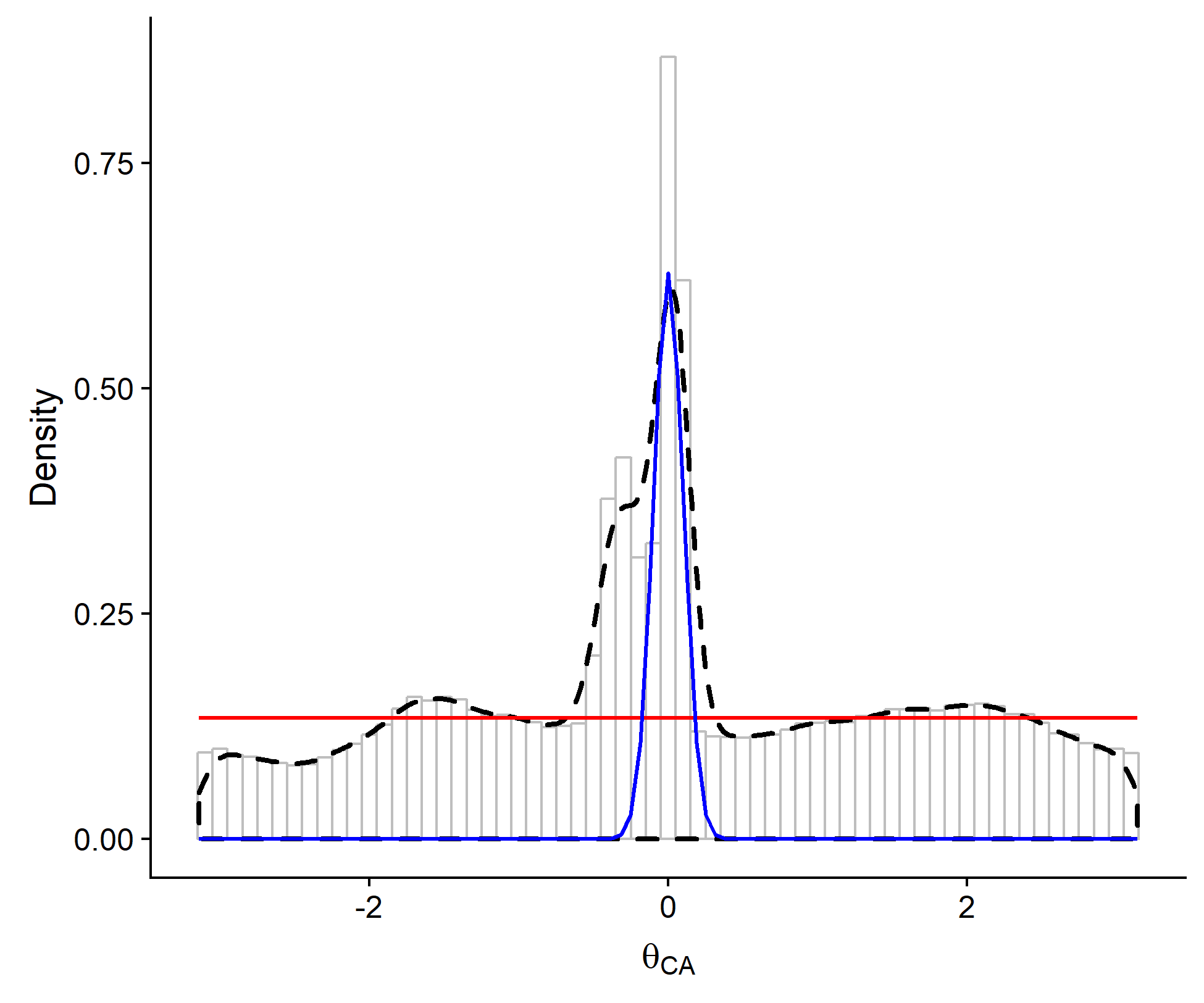}
			\caption{}
			\label{fig:dist.ca}
		\end{subfigure}
		\caption{Density Distributions. Red lines represent distributions associated with the MSH region, where as Blue lines represent those associated with the MSP.}
		\label{fig:distributions}
	\end{figure}
			
	The probability density distributions of the 395,458 values for each variable in our data set are generally bi-modal. Therefore, we can characterize the signals from either field by attributing their influence as a mixture of two normal distributions, representing the contributions from the magnetosheath (MSH) and magnetosphere (MSP) fields. When the satellite is in the MSH, data is normally distributed around $\mu_{MSH}$, with some standard error $\sigma_{MSH}$. In the MSP it is the same, with $\mu_{MSP}$ and $\sigma_{MSP}$. This allows us to, for any data point, establish the likelihood that the satellite is in either the MSH or MSP, based solely on the probability that a data value comes from either normal distribution.
	
	This is extremely helpful in identifying the MP region, the boundary where the influences of both fields blend. Data points with equal, but low likelihood of being in either field will have a high likelihood of belonging to the MP region. We introduce a mixture model that takes the prior probability density functions, or normal distributions, to estimate a mixing proportion $\lambda$—the likelihood of being in either field (Eq. ~\ref{eq:mixture}).
	
	Most of the variables we will use in our model can be estimated this way. One variable, the Clock Angle ($\theta_{CA}$, Figure \ref{fig:dist.ca}), cannot be estimated as a mixture of two normal distributions based on its inherent qualities. Any time the satellite is in the MSP, the $\theta_{CA}$ tends to remain close to 0 degrees. When the satellite is in the MSH region, $\theta_{CA}$ can rotate anywhere from $\pi$ to $-\pi$. This makes the $\theta_{CA}$ a mixture of normal and uniform distributions, as in Eq. \ref{eq:mixture:ca}.

	\begin{align}
		\label{eq:mixture}
		x \sim \lambda \, Normal(\mu_{MSH}, \sigma_{MSH}) + (1 - \lambda) \, Normal(\mu_{MSP}, \sigma_{MSP}) \\
		\label{eq:mixture:ca}
		x_{\theta_{CA}} \sim \lambda \, Uniform(-\pi, \pi) + (1 - \lambda) \, Normal(\mu_{MSP}, \sigma_{MSP})
	\end{align}
	
	In both Eq. \ref{eq:mixture} and Eq. \ref{eq:mixture:ca}, the variable being estimated by Stan is the $\lambda$ mixture ratio. When this ratio is close to 0, the likelihood that the satellite is in the MSP is very small, but it is very likely to be in the MSH. The reverse is true when the ratio is near 1, but more importantly, when the ratio is near 0.5 there is equal likelihood that the satellite is in either the MSH or MSP. It can be concluded that there is a high likelihood of being in the MP boundary layer when $\lambda$ mixing ratios are close to 0.5.
	
	Prior distribution $\mu$ and $\sigma$ values, used in the mixutre model, were generated using the normalmixEM method of the mixtools 1.0.4  R library, with k = 2 mixtures. This produced estimations using expectation maximization. For $B_t$ and $\theta_{CA}$, the $\mu$ and $\sigma$ values were adjusted manually for better predictive performance. The resulting values, represented by the red and blue lines in Figure \ref{fig:distributions}, were passed with the data to our model. All other $\alpha$, $\sigma$, $\lambda$, and $\beta$ values in equations \ref{eq:mixture}, ~\ref{eq:mixture:ca}, ~\ref{eq:regression}, and ~\ref{eq:linear} are estimated internally by stan.
	
	\subsection{Auto Regression Analysis}
	\begin{figure}
		\centering
		\begin{subfigure}[b]{0.45\textwidth}
			\includegraphics[width=\textwidth]{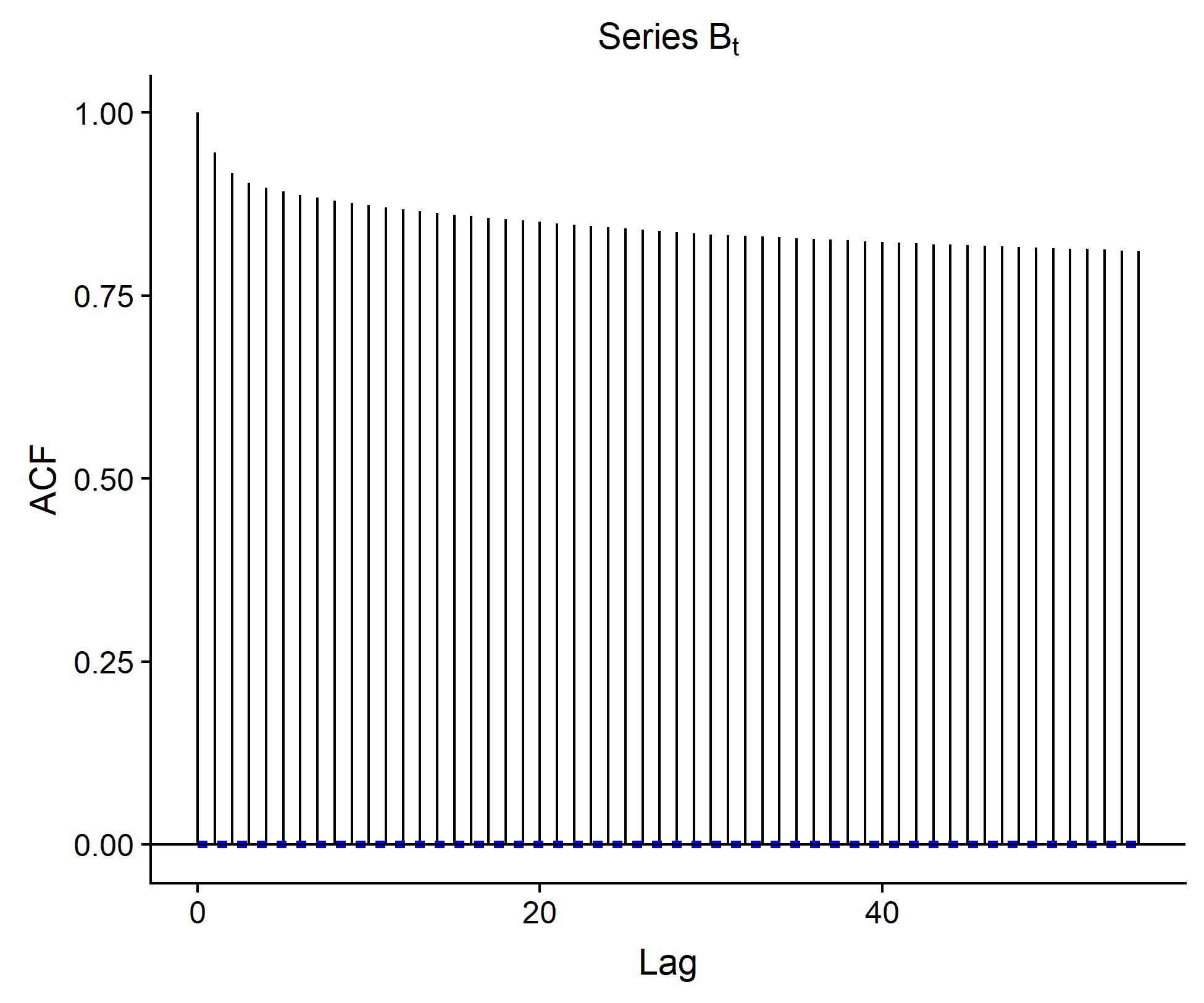}
			\caption{}
			\label{fig:bt.acf}
		\end{subfigure}
		\begin{subfigure}[b]{0.45\textwidth}
			\includegraphics[width=\textwidth]{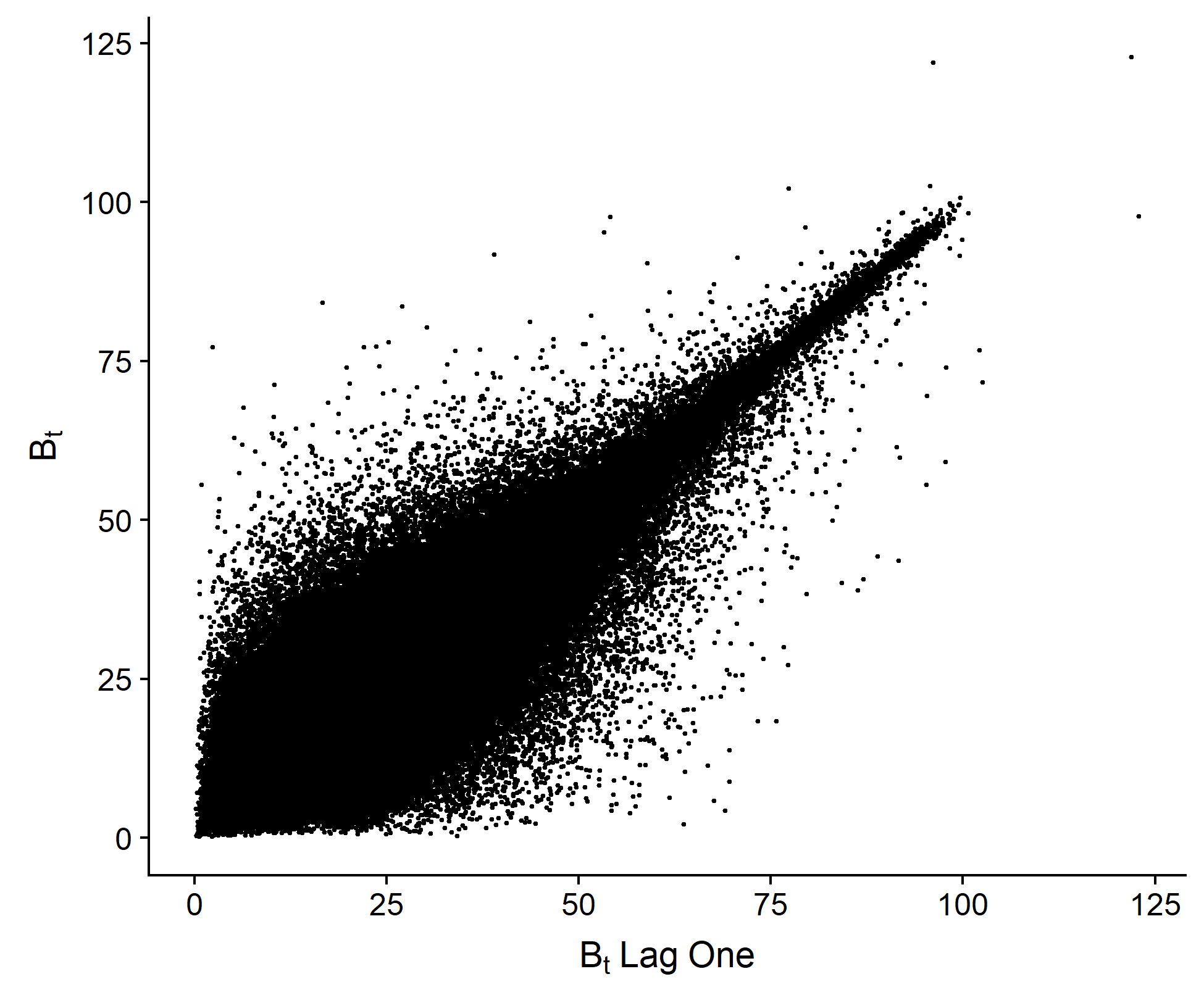}
			\caption{}
			\label{fig:bt.lagone}
		\end{subfigure}
		\begin{subfigure}[b]{0.45\textwidth}
			\includegraphics[width=\textwidth]{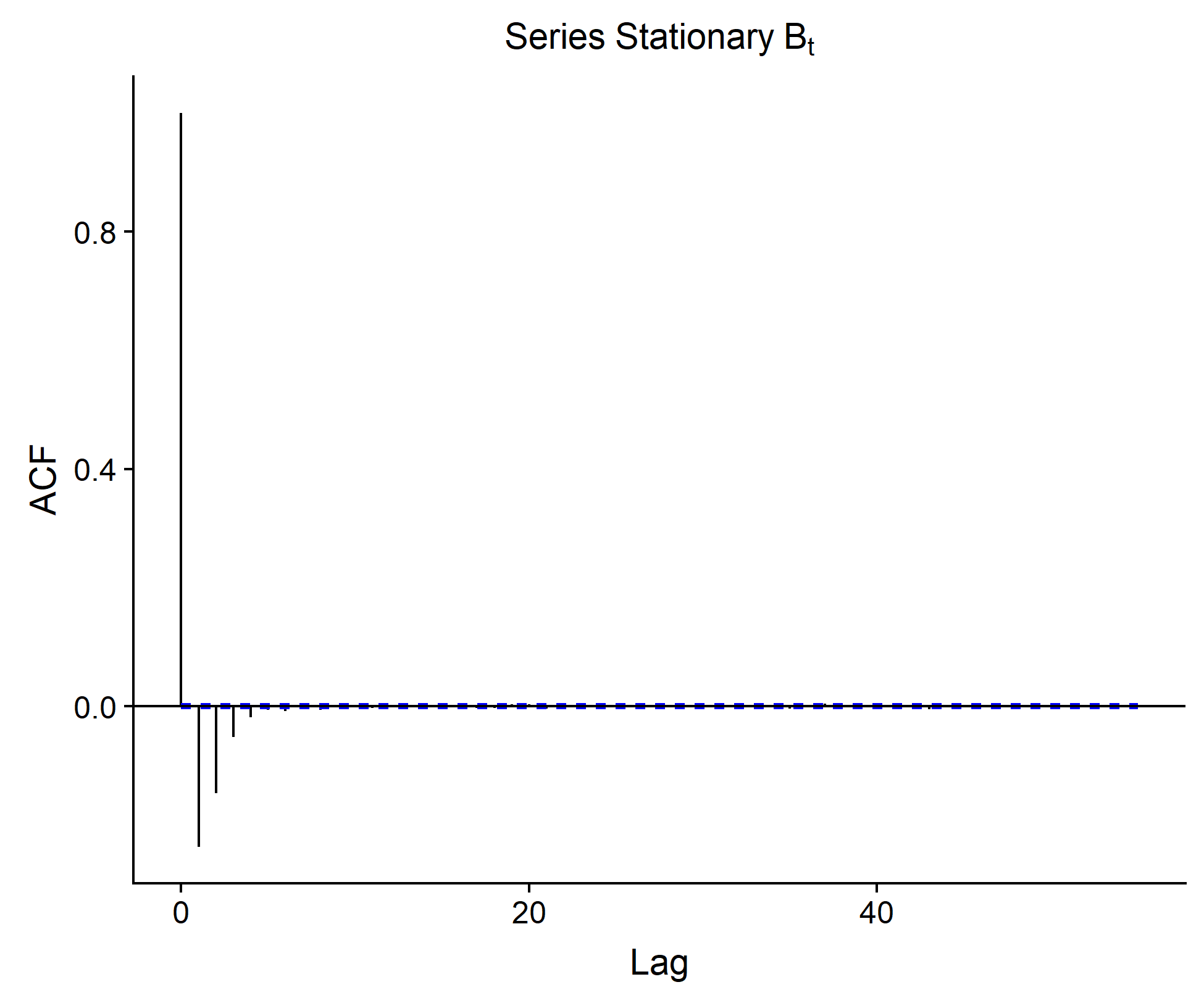}
			\caption{}
			\label{fig:bt.stationary}
		\end{subfigure}
		\begin{subfigure}[b]{0.45\textwidth}
			\includegraphics[width=\textwidth]{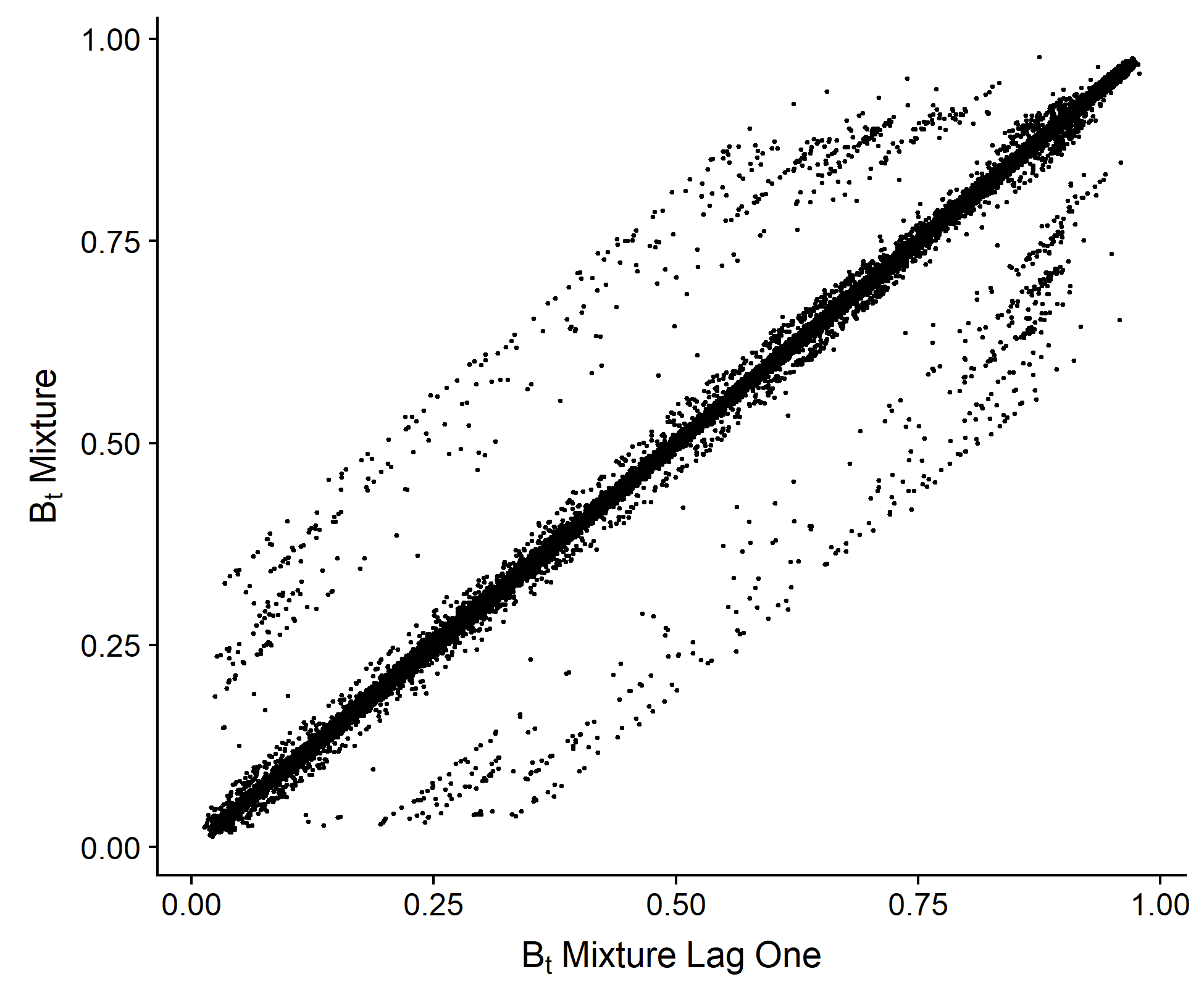}
			\caption{}
			\label{fig:bt.mix.lagone}
		\end{subfigure}
		\caption{Auto Regression Analysis. Plot A shows auto correlation values of $B_t$ at different lag times. C shows the same ACF plot, with $B_t$ stationary. Plots B and D show $B_t$ and the resulting $\lambda_{B_t}$ mixing ratios plotted against their lagged values, times t on the y axis and times t + 1 on the x axis. Additional plots for other variables are included in the appendix.}
		\label{fig:regressionBt}
	\end{figure}

	Intuitively, points in this data set are not independent from one another. As the satellite moves through space, its location is directly dependent on where it was before. Each variable exhibits non-stationarity, where the probability distribution changes based on which field the satellite is in or transitioning to at a given time. The data behaves as a Gaussian random walk, with each time point normally distributed around that of the last with a varying mean and standard deviation. This is verified by analyzing the auto correlation of each variable at different lag points. Figure \ref{fig:bt.acf} illustrates the non-stationarity of the variables ($B_t$ for example), and the scatter plots show the linear relationship of each time point against that of the next (Figure \ref{fig:bt.lagone}), at a lag of one. Further, once the data is made stationary by generating the difference at lag one, we can reanalyze the correlation of each time point again (Figure \ref{fig:bt.stationary}). Since there is no further relationship (Figure \ref{fig:bt.mix.lagone}), we conclude that lag one is appropriate for this model \cite{Hyndman2018}.
	
	\begin{align}
		\label{eq:regression}
		\lambda_t \sim Normal(\lambda_{t-1}, \sigma)
	\end{align}

	Therefore, an auto-regressive component is added to our model to account for this aspect of the data, using Eq.~\ref{eq:regression}. Each mixture ratio $\lambda$ at any time point is normally distributed around that of the last. Because the same behavior is exhibited in the resulting mixing ratio as seen in Figure \ref{fig:regressionBt}, we apply the auto regression to the mixing ratio $\lambda$. This has the effect of smoothing out $\lambda$, reducing the noise generated by the latent variable of time.
	
	\subsection{Linear Regression}	
	In order to retrieve a meaningful prediction, we regress on all resulting $\lambda$ values. This produces a number we can use to generate selections. The SITL provides a priority value, "Figure of Merit" (FOM), that can be used to train and test against. During January-February 2017, the SITLS attributed FOM values of 100 or more to MP regions. Eq.~\ref{eq:linear} models this priority variable using the $\lambda$ mixing ratios for each variable. The resulting priority predictions of 100 or more are classified as selections, to match what a SITL might assign as originating from the MP boundary layer.
	
	\begin{align}
		\label{eq:linear}
		\begin{split}
			Priority \sim & \alpha + \beta_{FGM_{Bt}} \, \lambda_{FGM_{Bt}} + \beta_{DIS_N} \, \lambda_{DIS_N} + \\ 
			& \beta_{DIS_T} \, \lambda_{DIS_T} + \beta_{\theta_{CA}} \, \lambda_{\theta_{CA}}
		\end{split}
	\end{align}

	\section{Evaluation and Results}
	We evaluated our model by comparing its performance against a basic tree boosting model. In addition to the true-positive rate, we evaluate with false-positive, miss-classification, and f-score. The true-positive rate reflects each model’s ability to select points in the MP, as noted by the SITL, while the true-negative rate reflects each model’s ability to ignore areas that the SITL did not note as being in the MP. Miss-classification shows how many points mismatched overall, and f-score (or harmonic mean) gives a holistic view of how balanced predictions are when compared to selections. Because SITL selections do not always completely encompass the MP region, and are somewhat subjective for each scientist, the SITL selections are considered a baseline for evaluation. Therefore, each evaluation metric on its own will not be a perfect indicator of the performance of models with this data set and goal. The null error performance is provided as a baseline, indicating what would happen if the dominant class were predicted 100\% of the time (in this case, no MP selections).
	
	For predictions using Gradient Boosting Machines (gbm), data was classified as selected or not for any FOM priority value of 100 or greater. Both models used the same four variables as well as the same test and training data set. The 'boosting' model was run using R’s gbm library with a maximum tree depth of 4, a maximum trees of 1000, and a Bernoulli distribution (giving binary 0/1 predictions using logistic regression).
	
	\begin{table}
		\begin{center}
			\begin{tabular}{|c|c|c|c|c|}
				\hline
				 & Miss-class. & True-positive & True-Negative & F-Score \\
				\hline
				Our Model & 13.16\% & 31.35\% & 93.02\% & 32.31\% \\
				Boosting & 12.74\% & 14.10\% & 95.40\% & 18.14\% \\
				Null Error & 10.02\% & 00.00\% & 100.00\% & 0.00\% \\
				\hline
			\end{tabular}
		\end{center}
		\caption{Evaluation results. Plots for the two other obits present in test data are included in the appendix.}
		\label{tab:results}
	\end{table}
	
\begin{figure}
		\centering
		\includegraphics[width=\linewidth]{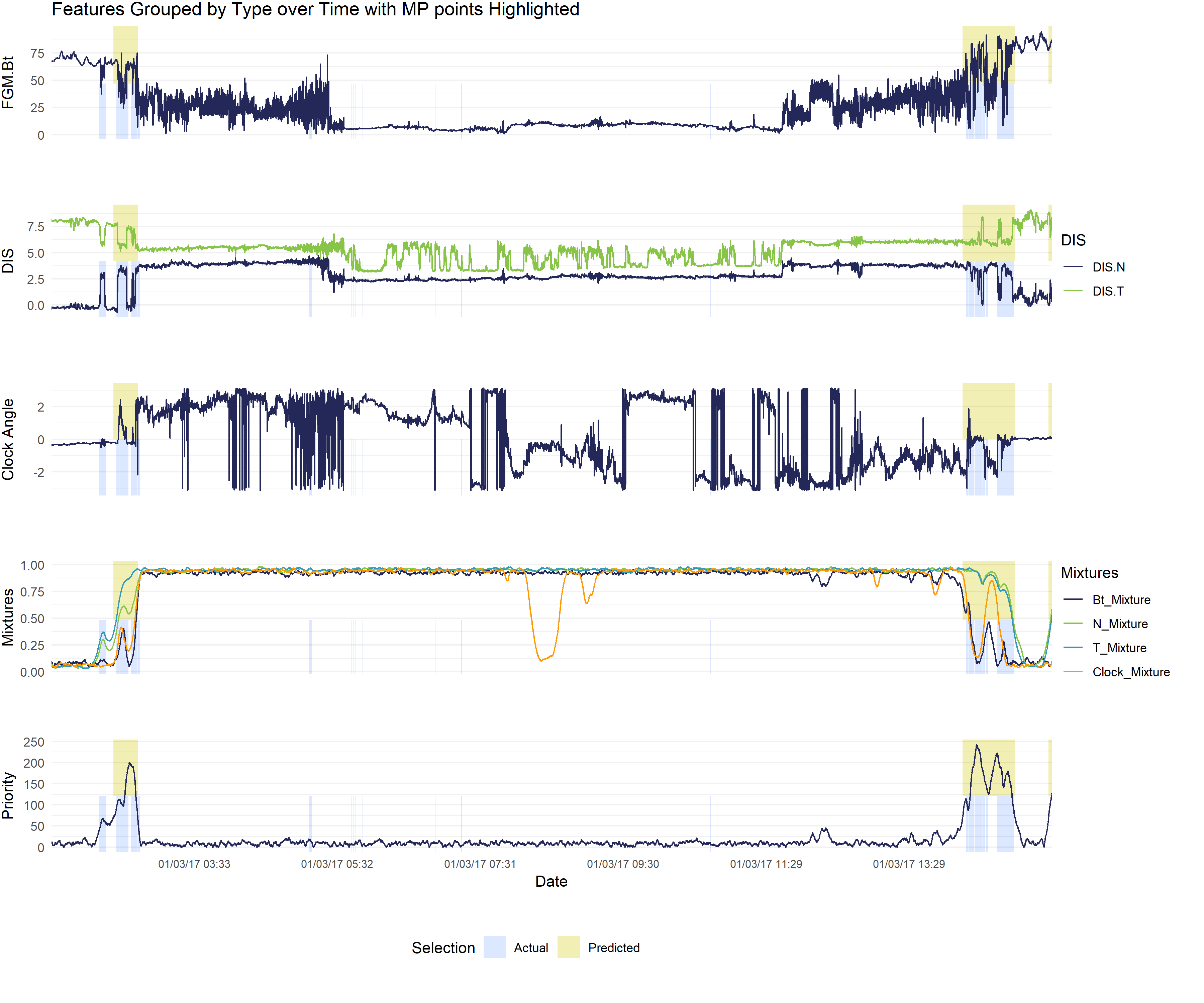}
		\caption{Resulting mixture ratios and predicted priorities for a single orbit, out of three test orbits.}
		\label{fig:results}
	\end{figure}

	Our model performs better or comparable on all four of the evaluation metrics outlined. Of note is our model's ability to predict two times as many MP regions, as identified by the SITL. This is without additional over selection and a competitively low miss-classification rate. We can also see, visually, that we are selecting regions that tend to align with shifts in values, as visible in Figure \ref{fig:results}.
	
	\section{Conclusions and Discussion}
	Our model performs exceptionally well based on evaluation metrics and visual inspection of its resulting selections. Moreover, through this process we have gathered meaningful information about the behavior of this data over time, the distribution of data, as well as the contribution of each feature variable we used to identify the MP boundary layer.
	
	The resulting beta values from the linear regression component indicate which mixing ratio contributes the most to accurate predictions. Because of the high evaluation scores of our model we can say that, of the four variables used ($B_t$, $n_i$, $T_i$, $\theta_{CA}$), the $B_t$ values tend to be the most indicative of being in the MP region. $T_{i}$ also contributes meaningfully to predictions.
	
	As mentioned before, while MP selections from the SITL are not incorrect, they may not completely encompass the entire MP region because of bandwidth constraints. Therefore, SITL selections tend to under select the MP region, which makes it an imperfect gold standard for this particular goal. We are still working on developing a way to more appropriately evaluate performance using SITL priority data.
	
	In addition to developing a more robust evaluation metric, we would like to increase our model’s performance by adding more variables, trying different ways of combining $\lambda$ mixing ratios, and fine-tuning hyper parameters. There are several variables that we were not able to incorporate into our model because of time and processing constraints. These variables could increase our ability to accurately select more MP boundary layers. Additionally, while linear regression allows us to see what mixture ratios are more useful to predictions, this function might not be the best fit with our goals. Processing time also limited our ability to do cross validation in order to tune model hyper parameters, such as selection threshold. Our naïve approach of selecting a threshold of 100, following what the SITL might give MP regions, works moderately well on most data but we could perform better with tuning.
	
	Our model selects twice as many MP regions than the comparative model, allowing scientists to use our selections to study the MP boundary layer and the reconnection events therein. Additionally, our model lays the foundation for machine learning’s application to MMS data and has huge potential to foster new discoveries to reconnection and the interaction between the solar wind and planetary magnetospheres. 
	
	\break
	
	\bibliography{references}{}
	\bibliographystyle{plain}
	
	\section{Appendix}
	\begin{figure}[H]
		\begin{center}
			\begin{tabular}{cc}
				\includegraphics[width=0.5\linewidth]{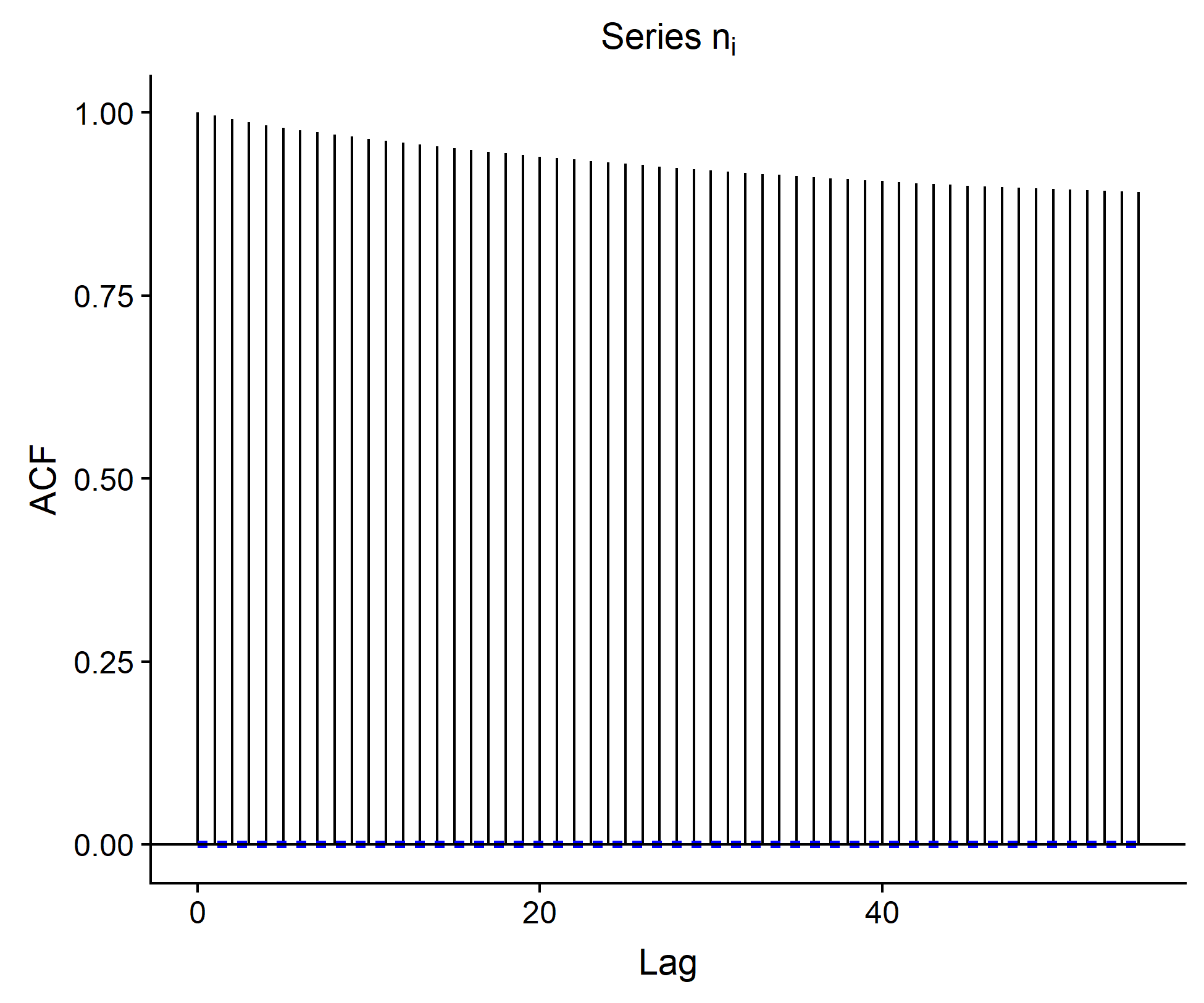} &
				\includegraphics[width=0.5\linewidth]{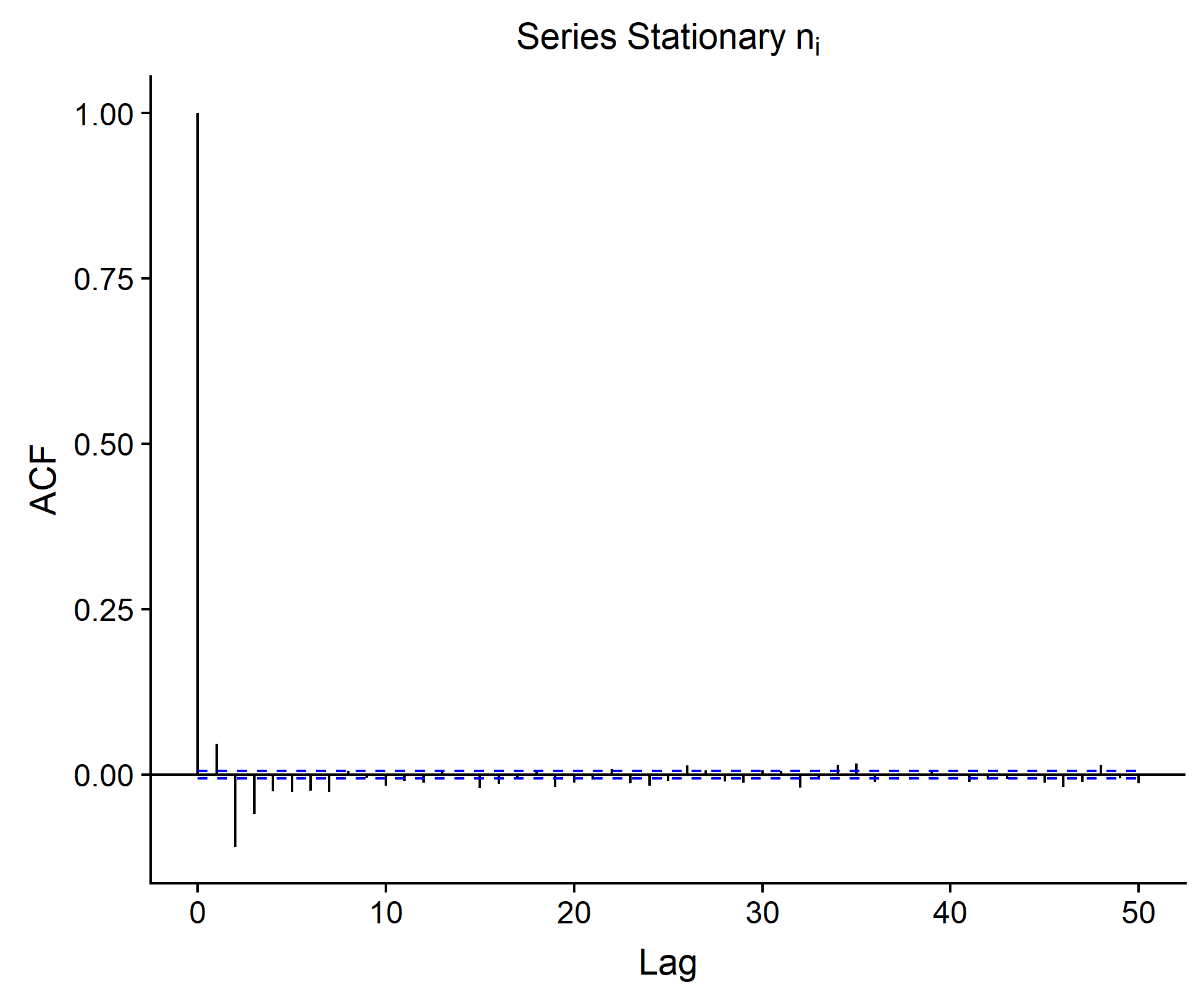} \\
				\includegraphics[width=0.5\linewidth]{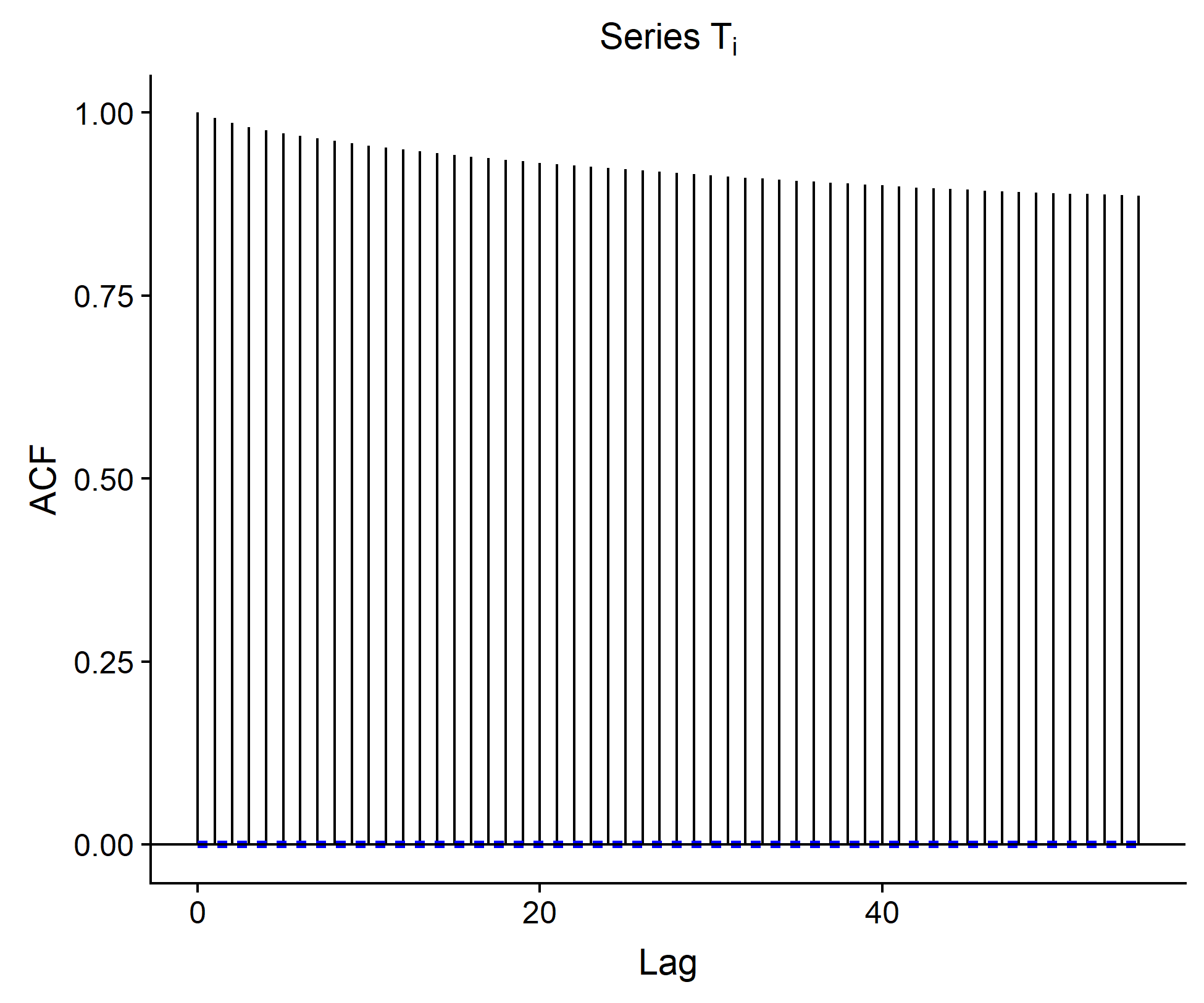} &
				\includegraphics[width=0.5\linewidth]{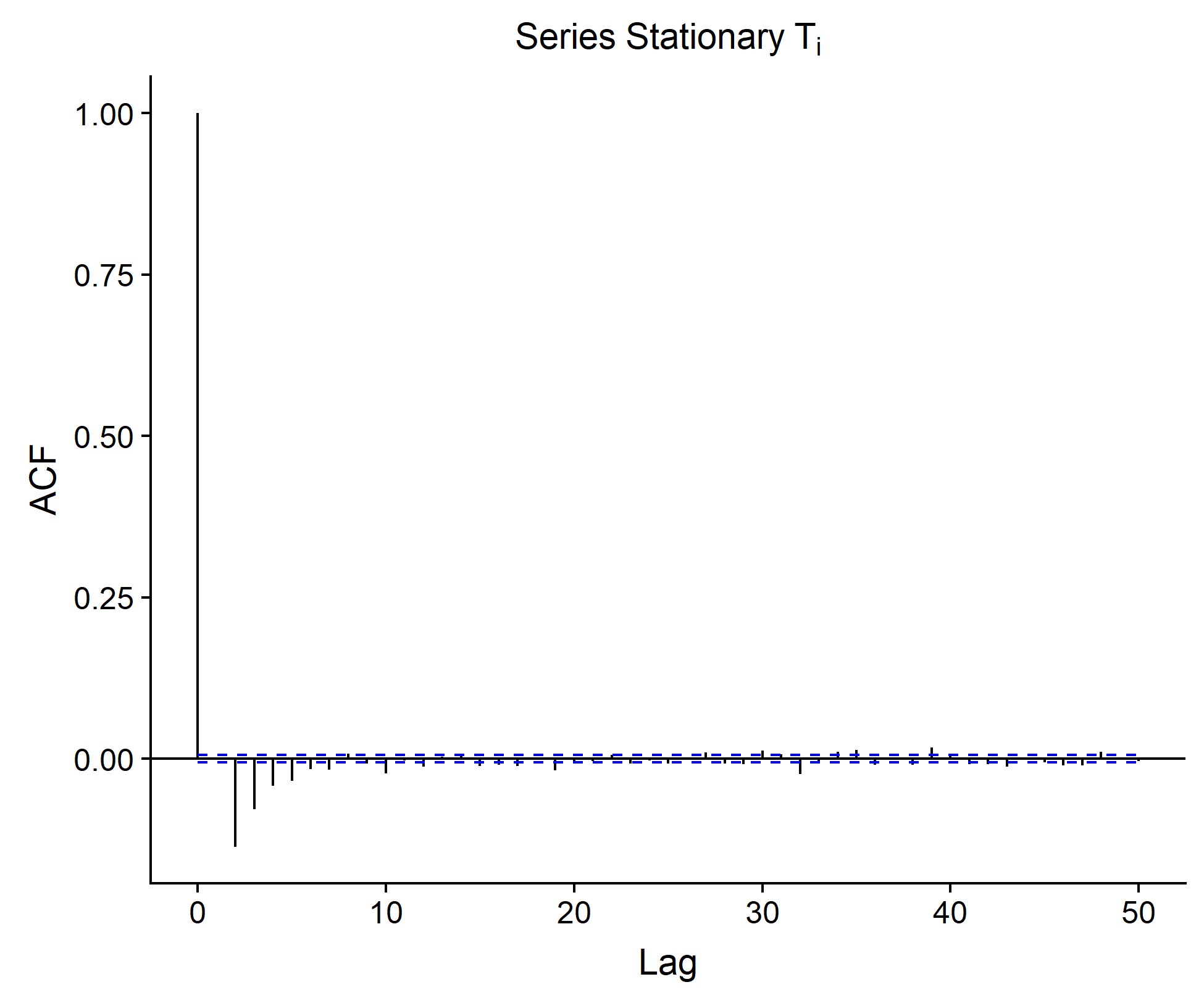} \\
				\includegraphics[width=0.5\linewidth]{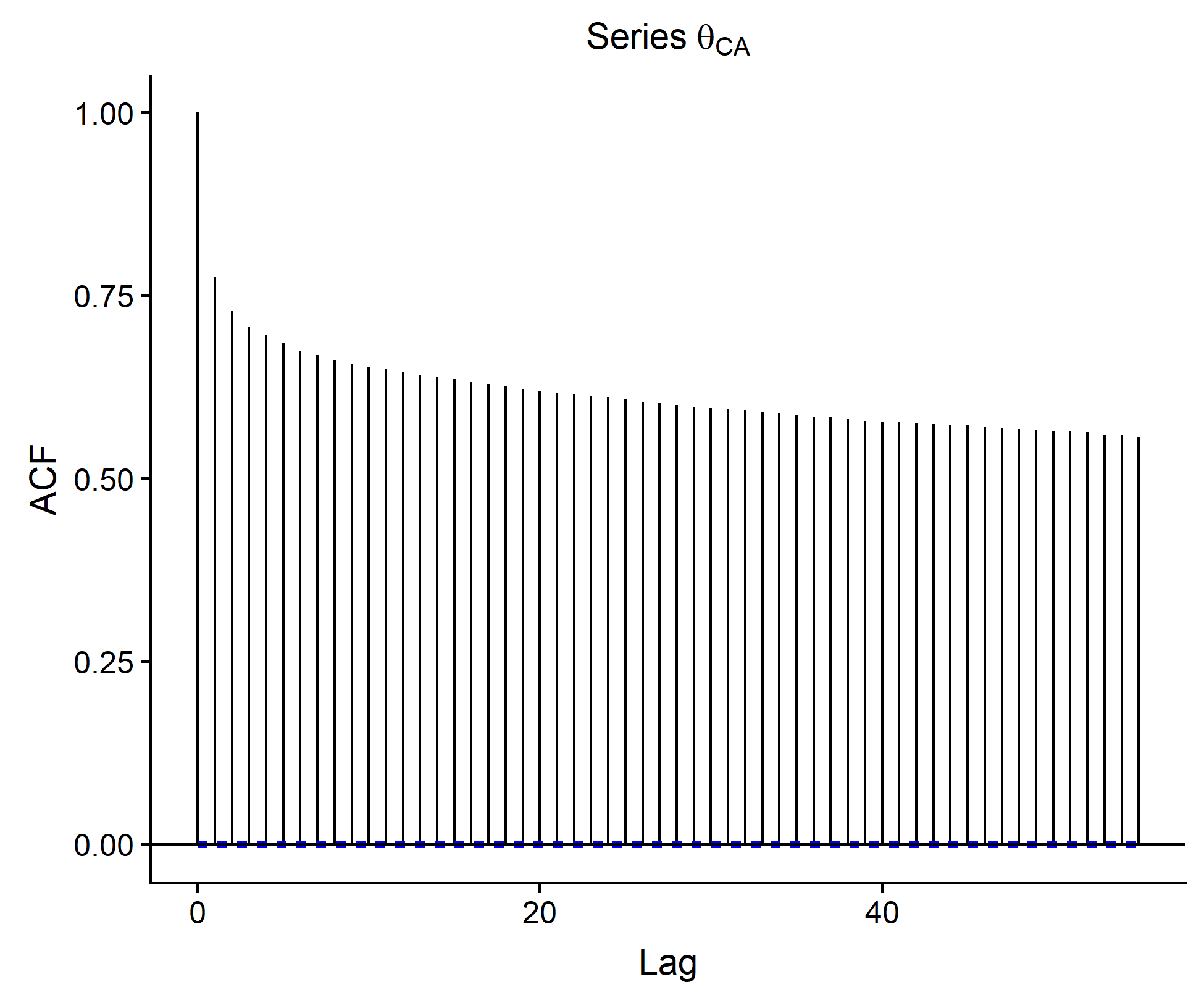} &
				\includegraphics[width=0.5\linewidth]{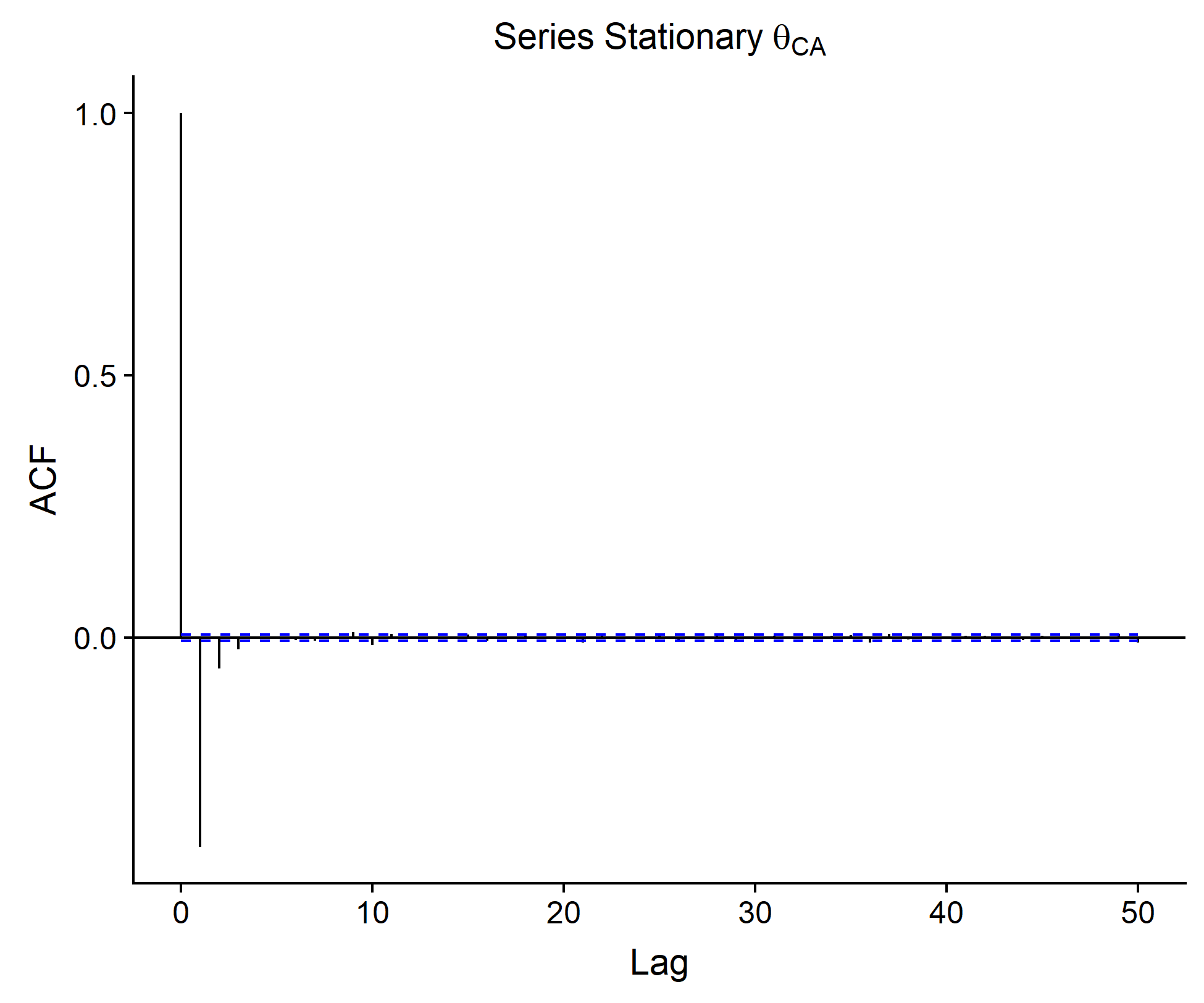}
			\end{tabular}
		\end{center}
		\caption{Auto Correlation Plots of raw variables and stationary variables.}
		\label{fig:acfs}
	\end{figure}

	\begin{figure}[H]
		\begin{center}
			\begin{tabular}{cc}
				\includegraphics[width=0.5\linewidth]{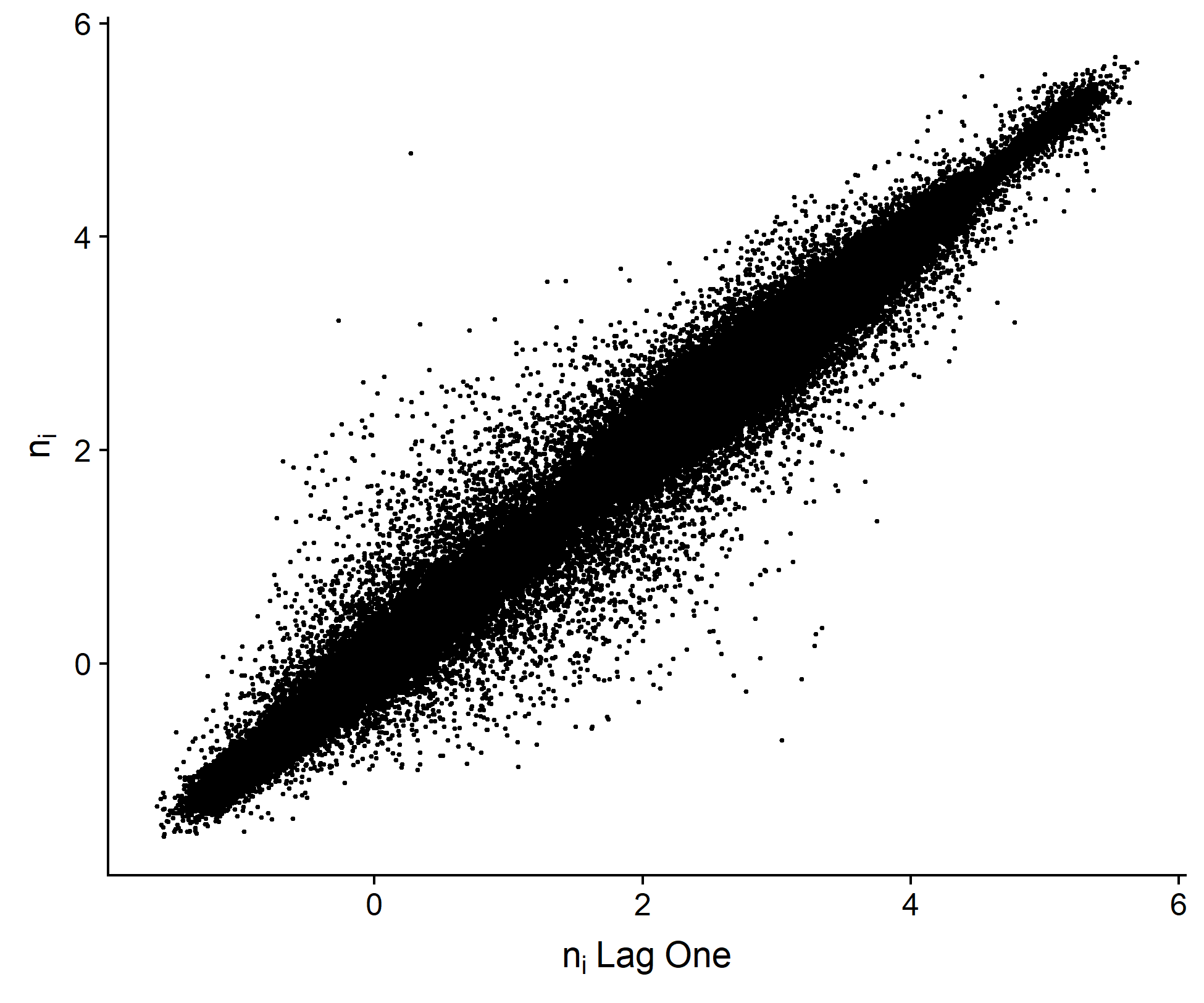} &
				\includegraphics[width=0.5\linewidth]{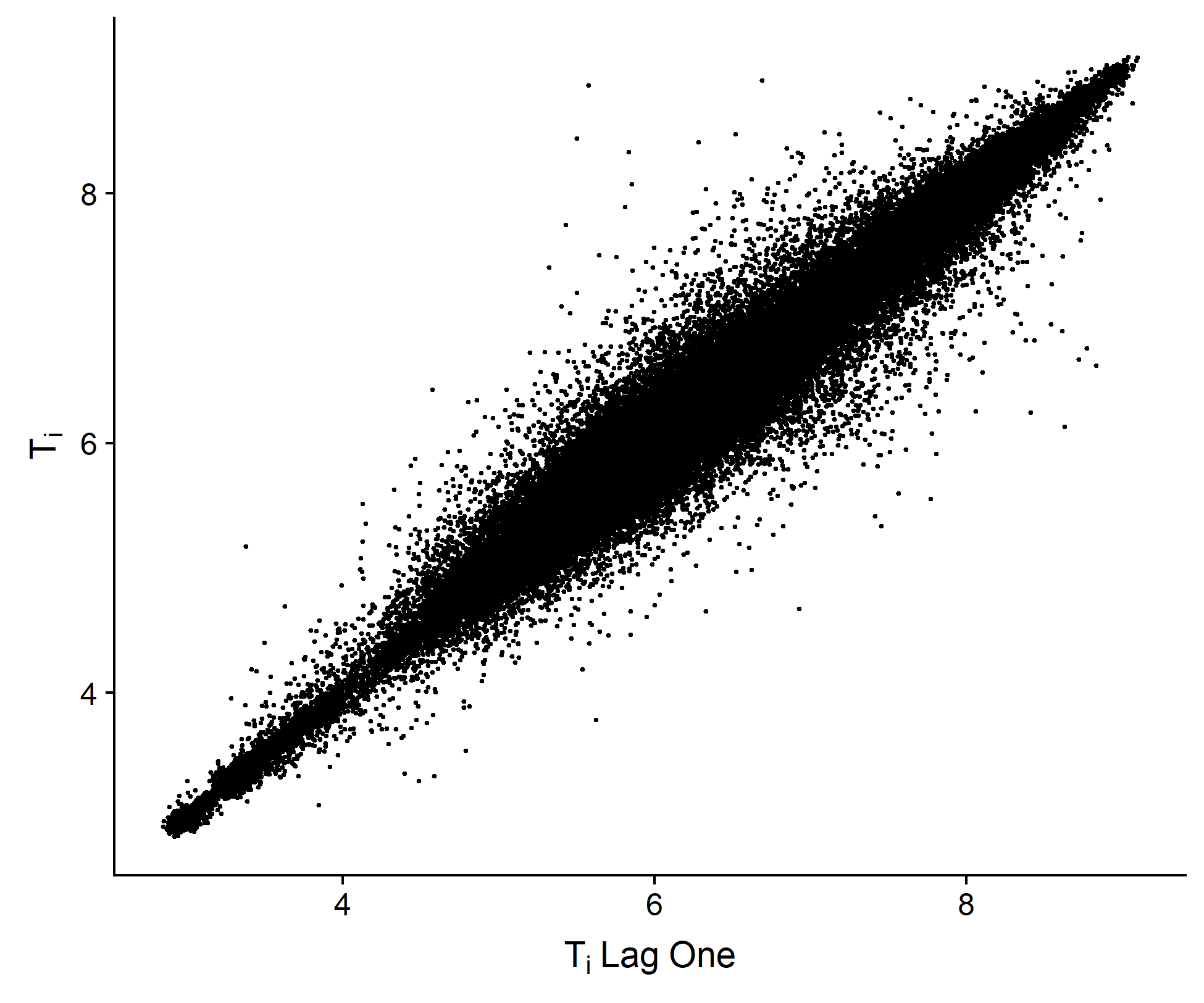} \\
				\includegraphics[width=0.5\linewidth]{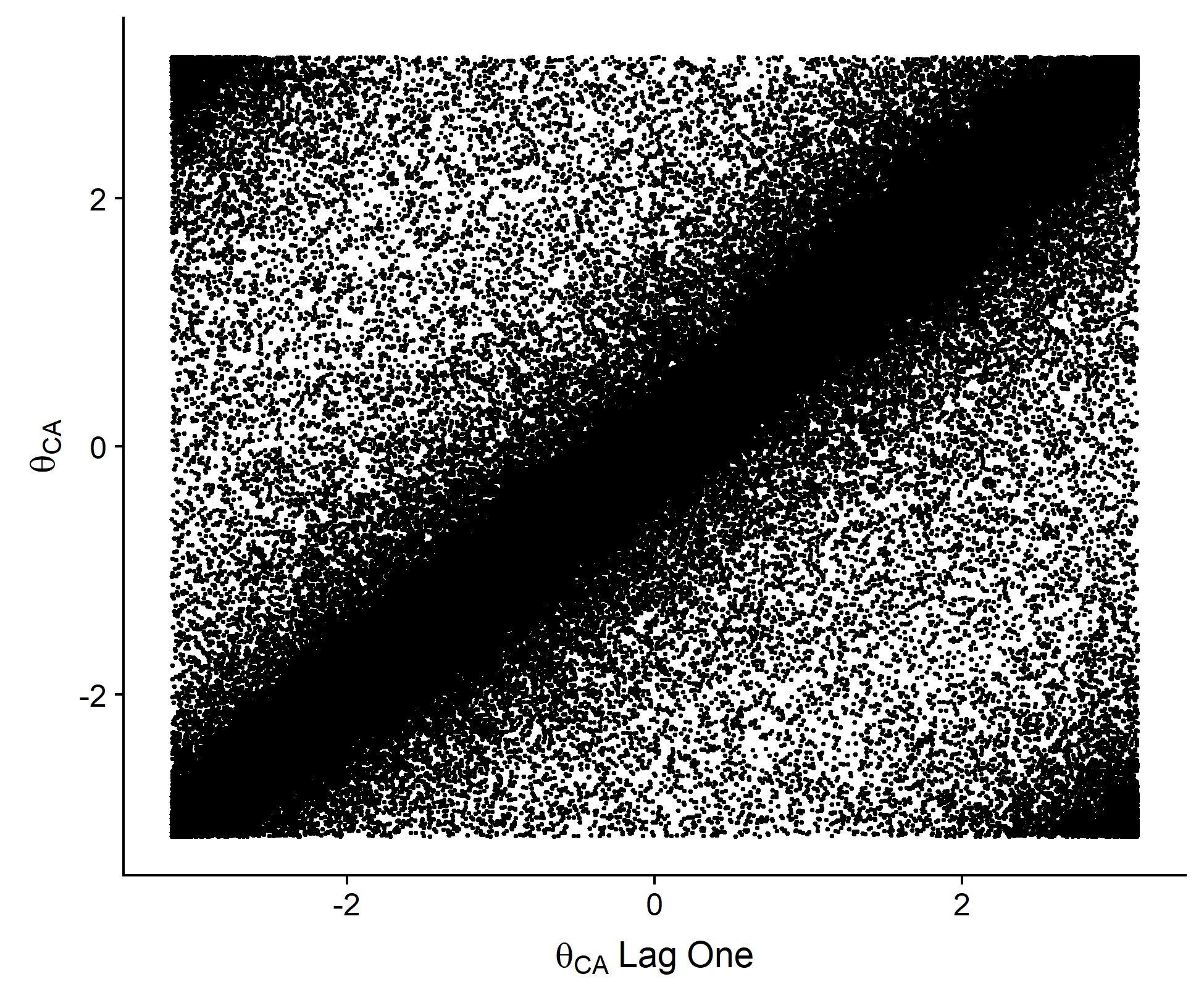}
			\end{tabular}
		\end{center}
		\caption{Lag one variable plots.}
		\label{fig:lagone}
	\end{figure}

	\begin{figure}[H]
		\centering
		\includegraphics[width=\linewidth]{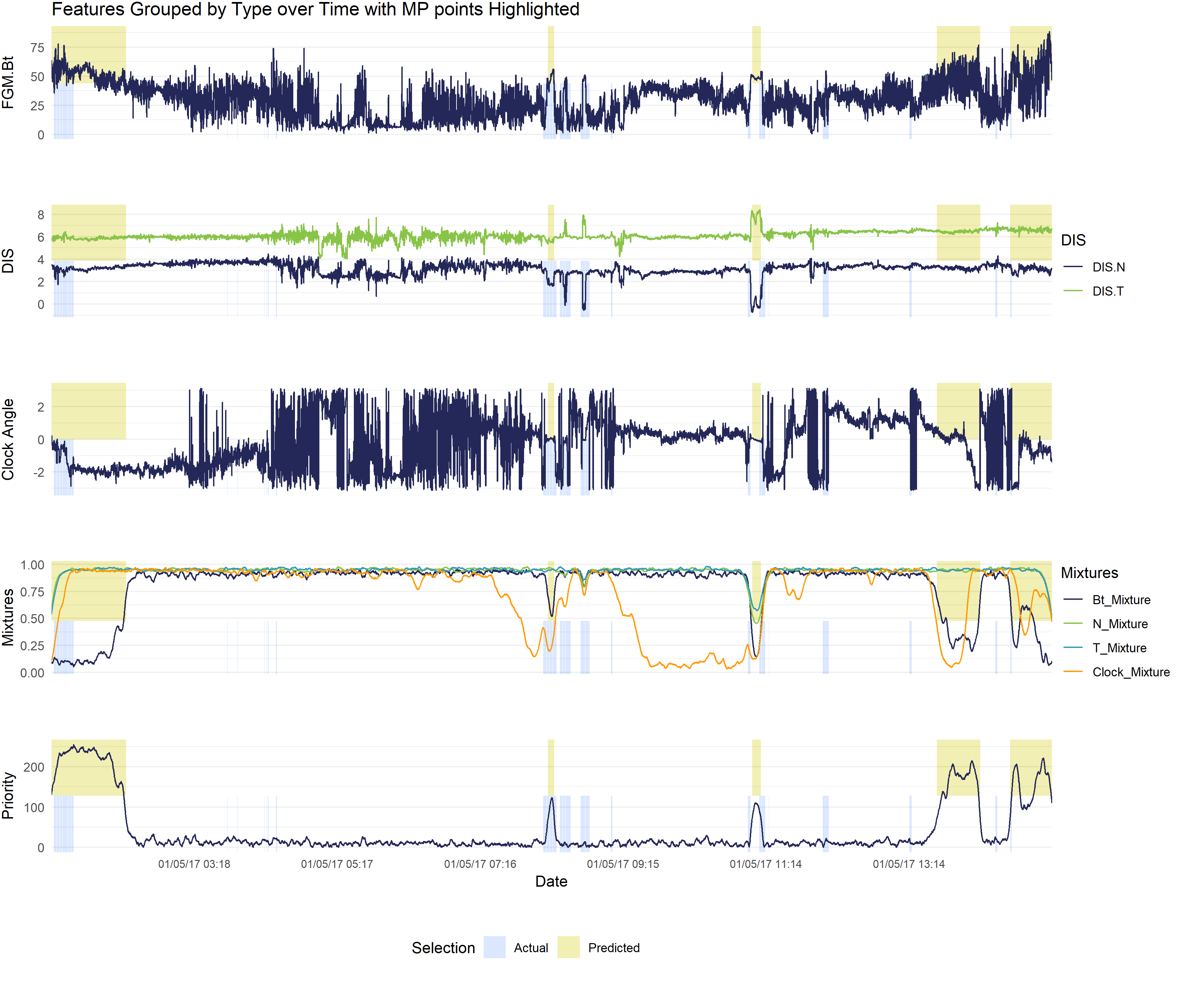}
		\caption{Test data predictions from model.}
		\label{fig:testPlots1}
	\end{figure}

	\begin{figure}[H]
		\centering
		\includegraphics[width=\linewidth]{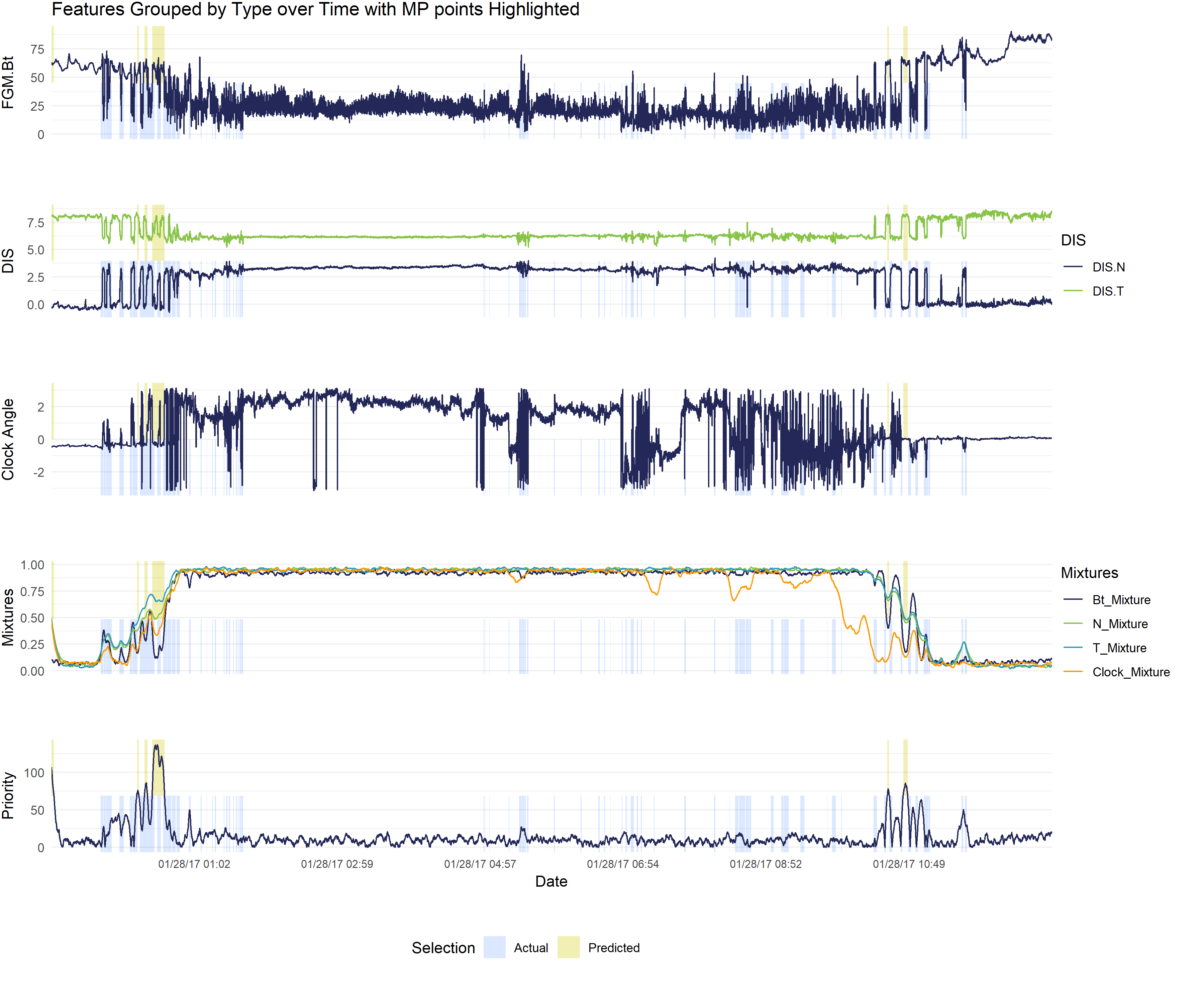}
		\caption{Test data predictions from model.}
		\label{fig:testPlots2}
	\end{figure}

	\break

	\begin{lstlisting}[style=stan, caption={Stan Training Model}, label={code:training}]
data{
	int numsteps;
	
	// Mix Prior Vectors: mu[MSH,MSP], sigma[MSH,MSP], theta
	vector[5] N_mix;
	vector[5] T_mix;
	vector[5] Bt_mix;
	
	// mu[MSH], sigma[MSH], Min[MSP], max[MSP]
	vector[4] Clock_mix;
	
	vector[numsteps] By;
	vector[numsteps] Bz;
	vector[numsteps] Bt;
	vector[numsteps] N;
	vector[numsteps] T_perp;
	vector[numsteps] T_para;
	vector<lower=0, upper=255>[numsteps] Priority;
}
transformed data{
	vector[numsteps] N_log;
	vector[numsteps] T_log;
	vector[numsteps] Clock_Angle;
	
	for (i in 1:numsteps){
		N_log[i] = log(N[i]);
		T_log[i] = log((T_para[i] + 2 * T_perp[i]) / 3);
		Clock_Angle[i] = atan2(By[i], Bz[i]);
	}
}
parameters{
	// Mixture Model
	vector<lower=0, upper=1>[numsteps] Bt_Mixture;
	vector<lower=0, upper=1>[numsteps] N_Mixture;
	vector<lower=0, upper=1>[numsteps] T_Mixture;
	vector<lower=0, upper=1>[numsteps] Clock_Mixture;
	real<lower=0> Clock_sigma;
	
	// Auto Regression
	real<lower=0> Bt_mix_sigma;
	real<lower=0> N_mix_sigma;
	real<lower=0> T_mix_sigma;
	real<lower=0> Clock_mix_sigma;
	
	// Linear Regression
	real mixture_alpha;
	real<lower=0> mixture_sigma;
	real Bt_beta;
	real N_beta;
	real T_beta;
	real Clock_beta;
}
model{
	// Mixture model
	for (n in 1:numsteps){
		target += log_mix(Bt_Mixture[n],
		normal_lpdf(Bt[n] | Bt_mix[1], Bt_mix[3]),
		normal_lpdf(Bt[n] | Bt_mix[2], Bt_mix[4]));
		
		target += log_mix(N_Mixture[n],
		normal_lpdf(N_log[n] | N_mix[1], N_mix[3]),
		normal_lpdf(N_log[n] | N_mix[2], N_mix[4]));
		
		target += log_mix(T_Mixture[n],
		normal_lpdf(T_log[n] | T_mix[1], T_mix[3]),
		normal_lpdf(T_log[n] | T_mix[2], T_mix[4]));
		
		target += log_mix(Clock_Mixture[n],
		uniform_lpdf(Clock_Angle[n] | Clock_mix[3], Clock_mix[4]),
		normal_lpdf(Clock_Angle[n] | Clock_mix[1], Clock_sigma));
	}

	// Auro-Regression
	for (n in 2:numsteps){
		Bt_Mixture[n] ~ normal(Bt_Mixture[n-1], Bt_mix_sigma);
		N_Mixture[n] ~ normal(N_Mixture[n-1], N_mix_sigma);
		T_Mixture[n] ~ normal(T_Mixture[n-1], T_mix_sigma);
		Clock_Mixture[n] ~ normal(Clock_Mixture[n-1], Clock_mix_sigma);
	}

	// Linear Regression
	Priority ~ normal(mixture_alpha + Bt_beta * Bt_Mixture +
	N_beta * N_Mixture + T_beta * T_Mixture +
	Clock_beta * Clock_Mixture, mixture_sigma);
}
	\end{lstlisting}
	\begin{lstlisting}[style=stan, caption={Stan Testing Model}, label={code:testing}]
data{
	int numsteps;
	
	// Mix Prior Vectors: mu[MSH,MSP], sigma[MSH,MSP], theta
	vector[5] N_mix;
	vector[5] T_mix;
	vector[5] Bt_mix;
	
	// mu[MSH], sigma[MSH], Min[MSP], max[MSP]
	vector[4] Clock_mix;
	real<lower=0> Clock_sigma;
	
	// Vector values
	vector[numsteps] By;
	vector[numsteps] Bz;
	vector[numsteps] Bt;
	vector[numsteps] N;
	vector[numsteps] T_perp;
	vector[numsteps] T_para;
	
	// Auto Regression
	real<lower=0> Bt_mix_sigma;
	real<lower=0> N_mix_sigma;
	real<lower=0> T_mix_sigma;
	real<lower=0> Clock_mix_sigma;
	
	// Linear Regression
	real mixture_alpha;
	real<lower=0> mixture_sigma;
	real Bt_beta;
	real N_beta;
	real T_beta;
	real Clock_beta;
}
transformed data{
	vector[numsteps] N_log;
	vector[numsteps] T_log;
	vector[numsteps] Clock_Angle;
	
	for (i in 1:numsteps){
		N_log[i] = log(N[i]);
		T_log[i] = log((T_para[i] + 2 * T_perp[i]) / 3);
		Clock_Angle[i] = atan2(By[i], Bz[i]);
	}
}
parameters{
	// Mixture Model
	vector<lower=0, upper=1>[numsteps] Bt_Mixture;
	vector<lower=0, upper=1>[numsteps] N_Mixture;
	vector<lower=0, upper=1>[numsteps] T_Mixture;
	vector<lower=0, upper=1>[numsteps] Clock_Mixture;
}
model{
	// Mixture model
	for (n in 1:numsteps){
		target += log_mix(Bt_Mixture[n],
		normal_lpdf(Bt[n] | Bt_mix[1], Bt_mix[3]),
		normal_lpdf(Bt[n] | Bt_mix[2], Bt_mix[4]));
		
		target += log_mix(N_Mixture[n],
		normal_lpdf(N_log[n] | N_mix[1], N_mix[3]),
		normal_lpdf(N_log[n] | N_mix[2], N_mix[4]));
		
		target += log_mix(T_Mixture[n],
		normal_lpdf(T_log[n] | T_mix[1], T_mix[3]),
		normal_lpdf(T_log[n] | T_mix[2], T_mix[4]));
		
		target += log_mix(Clock_Mixture[n],
		uniform_lpdf(Clock_Angle[n] | Clock_mix[3], Clock_mix[4]),
		normal_lpdf(Clock_Angle[n] | Clock_mix[1], Clock_sigma));
	}

	// Auro-Regression
	for (n in 2:numsteps){
		Bt_Mixture[n] ~ normal(Bt_Mixture[n-1], Bt_mix_sigma);
		N_Mixture[n] ~ normal(N_Mixture[n-1], N_mix_sigma);
		T_Mixture[n] ~ normal(T_Mixture[n-1], T_mix_sigma);
		Clock_Mixture[n] ~ normal(Clock_Mixture[n-1], Clock_mix_sigma);
	}
}
generated quantities {
	vector[numsteps] Priority;
	
	// Linear Regression
	for (i in 1:numsteps){
		Priority[i] = normal_rng(mixture_alpha + Bt_beta * Bt_Mixture[i] +
		N_beta * N_Mixture[i] + T_beta * T_Mixture[i] +
		Clock_beta * Clock_Mixture[i], mixture_sigma);
	}
}
	\end{lstlisting}
	Additional R files used to process data and run the model can be found at:
	https://github.com/srpiatt/MMS\_Magnetopause\_MixtureModel
\end{document}